\documentclass[nofootinbib,twocolumn,pra,letterpaper,longbibliography,,superscriptaddress]{revtex4-2}
\usepackage{multirow}
\usepackage{latexsym,amsmath,amssymb,amsfonts,graphicx,color,amsthm}
\newtheorem{theorem}{Theorem}
\usepackage{enumerate}
\usepackage{braket}
\usepackage{adjustbox}
\usepackage{footnote}
\usepackage[dvipsnames]{xcolor}
\usepackage{tipa}
\usepackage{textcomp}
\usepackage{fontenc}
\usepackage{mathrsfs}
\usepackage{color}
\usepackage{colortbl}
\usepackage{graphics,graphicx}
\usepackage{txfonts}
\usepackage{caption}
\usepackage{subcaption}
\usepackage{cancel}

\usepackage{amsmath}
\usepackage{amsfonts,amssymb,amstext,amscd,amsthm,bbold}
\usepackage{comment,float}
\newcommand{\tr}{\textcolor{red}}

\usepackage[colorlinks=true,linkcolor=blue,urlcolor=magenta,citecolor=magenta]{hyperref}

\definecolor{myurlcolor}{rgb}{0,0,0.7}

\usepackage{float}
\usepackage{hyperref}
\def\tr{\operatorname{tr}}

\newcommand{\pt}{\mathcal{P}\mathcal{T}}        
\newcommand{\G}{\mathcal{G}}
\newcommand{\h}{\mathcal{H}}
\renewcommand{\ket}[1]{\vert#1\rangle}

\renewcommand{\bra}[1]{\langle#1\vert}
\newcommand{\miniprod}[2]{\langle#1\vert#2\rangle}

\begin{document}
%\title{Discrete non-Hermitian dynamics}
\title{Reduced dynamics in quasi-Hermitian systems}
\author{Himanshu Badhani}
\email{himanshub@imsc.res.in, himanshubadhani@gmail.com}
\affiliation{The Institute of Mathematical Sciences, C. I. T. Campus, Taramani, Chennai 600113, India}
\affiliation{Homi Bhabha National Institute, Training School Complex, Anushakti Nagar, Mumbai 400094, India}
%\author{Subhashish Banerjee}
%\affiliation{Indian Institute of Technology Jodhpur, Jodhpur-342011, India}
%\email{subhashish@iitj.ac.in}
\author{C. M. Chandrashekar}
\email{chandru@imsc.res.in}
\affiliation{The Institute of Mathematical Sciences, C. I. T. Campus, Taramani, Chennai 600113, India}
\affiliation{Homi Bhabha National Institute, Training School Complex, Anushakti Nagar, Mumbai 400094, India}
\affiliation{Department of Instrumentation \& Applied Physics, Indian Institute of Science, Bengaluru 560012, India}
%===============================================================================================

\begin{abstract}
Evolutions under non-Hermitian Hamiltonians with unbroken $\pt$ symmetry can be considered unitary under appropriate choices of inner products, facilitated by the so-called metric operator.  While it is understood that the choice of the metric operator has no bearing on the description of the system, in this work, we show that this choice does dictate the entanglement structure of the system. We show that the partial trace of the Hermitized density matrix gives the correct representation of the reduced subsystem, and based on such operations, we elucidate the metric dependency of the reduced dynamics and consequently the observable dependence of the subsystem decomposition. We use a non-Hermitian $\mathcal{PT}$-symmetric quantum walk as a toy model to study this metric dependency, where we use the internal (coin state) as the subsystem of interest and study the coin-position entanglement and non-Markovianity of the coin dynamics.
\end{abstract}

%==================================================================================
\maketitle
\section{Introduction}
\noindent
Hamiltonians invariant under the simultaneous action of parity and time reversal operations ($\pt$) have eigenvalues in complex conjugate pairs. In certain parameter regimes, such Hamiltonians have real eigenvalues, even if they are not Hermitian\,\cite{Bender98, Bender02}. These Hamiltonians have been a topic of increasing interest in recent years\ \cite{hatano96,rudner09,Pes12,De23,Naom12} and can be simulated using balanced gain and loss mechanisms~\cite{cheng1,cheng2} and find use in quantum technologies~\cite{javid}. It turns out that there is  an equivalence between the non-unitary $\pt$-symmetric dynamics and the unitary dynamics \cite{Ali02, Ali03, bender2, ali02p1, rotter} that is established by introducing a metric operator which redefines the inner product. In fact, such a description is essential to avoid nonphysical results like the violation of the no-signaling principle\ \cite{Lee14}. However, the choice of the metric operator is not unique, and given a Hamiltonian, one can choose from an infinite class of metric operators. All choices of the metric operator are equivalent since the expectation values of the observables in all inner product spaces are the same if the state and the observables are transported to the different inner product spaces properly. This is the framework of pseudo-Hermitian quantum mechanics. In this work, we will study the intricacies of dealing with composite quantum systems under this framework. For composite quantum systems, the Hilbert space is assumed to have a tensor product structure, with each component of the tensor product representing the subsystem. In such Hilbert space structures, the partial trace operation is the unique, completely positive, and trace-preserving mapping to the Hilbert space of the subsystem. While dealing with a non-Hermitian $\pt$-symmetric Hamiltonian for a composite quantum system, a change in the inner product structure on the total Hilbert space need not preserve the tensor product structure. This leads to the problem of defining the subsystem in such a framework since the partial trace operation is not well-defined. To circumvent this issue, previous studies have focused on certain choices of metric operators to make the partial trace operation well defined, in particular using a metric in tensor product form\ \cite{moise,nori} where the Hilbert space retains the direct product structure, or in a direct sum form \cite{hb23} where the partial trace can be performed by exploiting the peculiar geometric structure of the state. However, there is no physical principle that dictates one choice of the metric operator over the other. Moreover, as we will show, not all Hamiltonians might allow such convenient choices of the metric operator.
\\\\
In this work, we show that the subsystems for states in such Hilbert spaces can be well defined through a partial trace on the Hermitized states, which are the unique, self-adjointness-preserving maps to the states in the Hilbert space with the conventional Euclidean inner product. This allows us to construct subsystems for all choices of the metric operators compatible with the Hamiltonian. This also leads us to the conclusion that the choice of the inner product can result in different properties of the subsystem, indicating that the subsystem decomposition is metric-dependent. The choice of the metric operator is determined by a set of observables used to probe the system\ \cite{ scholtz}. Previous works, e.g.\ \cite{paolo01}, have concluded that the subsystem's properties, like entanglement, which depend on the tensor product structure of the system, do depend on the choice of the observables. Our conclusions in this manuscript extend these observations to pseudo-Hermitian quantum mechanics.
\\\\
In section \ref{sec:intro}, we introduce the pseudo-Hermitian formalism with a focus on the Hermitization map and non-uniqueness of the metric operator. We present the arguments for partial trace through Hermitization in section\ \ref{sec:subsystems}, where we also observe the effect of the choice of the metric on the subsystem decomposition. In section\ \ref{sec:qw} we study the $\pt$-symmetric quantum walk model and treat it as a composite system. We show that this quantum walk operation does not allow a metric operator in tensor product form, and the interactions between the subsystems (its coin and position subspaces) depend on the choice of the metric.
%==================================================================================
\noindent
%==================================================================================
\section{Introduction to Pseudo-Hermitian quantum mechanics}\label{sec:intro}
While all Hermitian Hamiltonians have real eigenvalues, Hermiticity is not a necessary condition for the eigenvalues to be real. Consider the operators that commute with an anti-linear involution operator $\mathbb{U}$ \ \cite{Bender07} . An anti-linear involution satisfies the following two conditions $\mathbb{U}$ on the Hilbert space $\mathcal{H}$: 
\begin{equation}
\mathbb{U}a\phi=a^*\mathbb{U}\phi,\hspace{1cm}\text{where,}\hspace{1cm} \mathbb{U}^2=1,
\end{equation}
where $a\in \mathbb{C}$ and $\phi\in \mathcal{H}.$ If $H$ commutes with this anti-linear involution, that is $[\mathbb{U},H]=0$, the eigenvalues of $H$ come in complex conjugate pairs with equal multiplicity, while if it also shares the eigenvectors with $\mathbb{U}$, the corresponding eigenvalues are guaranteed to be real. One such example of an anti-unitary operation is the time reversal operation $\mathcal{T}$, which implements the transformation $t\to -t$, where $t$ is the time parameter. Given a linear operation, for instance, the parity reversal operation $\mathcal{P}$, which corresponds to the transformation $x\to -x$ where $x$ is a spatial coordinate, the $\pt$ operation is an anti-linear operation. An operation invariant under $\pt$ operation either has real eigenvalues or its eigenvalues are complex conjugate pairs with the same multiplicity. When the eigenvalues are real, it is called unbroken $\pt$ symmetry, contrasting it with the case of $\pt$-broken symmetry when the eigenvectors are not eigenvectors of the $\pt$-operation. 
\\\\ 
It was realized that diagonalizable operators $\mathcal{O}:\h\rightarrow \h$ that are invariant under an anti-linear involution satisfy a property called \textit{G-pseudo-Hermiticity}\ \cite{ali02p1} : for such operators there exists an invertible, Hermitian operator $G:\h\to\h$, such that $G^{-1}\mathcal{O}^\dagger G=\mathcal{O}$\ \cite{ali02p3}.  Therefore, every diagonalizable $\pt$-symmetric operator is also a $G$-pseudo-Hermitian operator.

On the other hand, if $\mathcal{O}$ is \textit{quasi-Hermitian}, it is related to a Hermitian operator $\mathcal{O}_\eta$ through a similarity transformation $\mathcal{O}=\eta^{-1}\mathcal{O}_\eta\eta$, for some $\eta$. One can see that $G$-pseudo-Hermiticity in certain cases can imply quasi-Hermiticity, particularly when the operator $G$ is also a positive-definite operator\ \ \cite{Ali10}. This is due to the fact that for such operators, there exists a unique square root which is also Hermitian and positive-definite, let us call it $\eta$, i.e. $G=\eta^2$. Then, the condition of pseudo-Hermiticity implies
\begin{equation}
\begin{aligned}
        \mathcal{O}^\dagger G &=G\mathcal{O}\\
        \Rightarrow \eta^{-1}\mathcal{O}^\dagger \eta &=\eta \mathcal{O}\eta^{-1}.
\end{aligned}
\end{equation}
Here $\mathcal{O}_\eta=\eta \mathcal{O} \eta^{-1}$ is the Hermitian operator\footnote{In the literature, the metric operator is generally denoted by $\eta$ and its positive square root by $\rho$. We find this notation unnecessarily confusing, hence the choice.}. It is also easy to see that every quasi-Hermitian operator is also pseudo-Hermitian, as given an operator $\eta$ satisfying $\mathcal{O}=\eta^{-1}\mathcal{O}_\eta\eta$\,, the positive-definite operator $G=\eta^\dagger \eta$  satisfies the relation $\mathcal{O}^\dagger G=G\mathcal{O}$. Therefore, quasi-Hermiticity is a sufficient condition for the operator to be diagonalizable with a real spectrum. It is also a necessary condition for the same\ \cite{Ali02}, since for such operators one can use their left-eigenvectors to construct the compatible positive-definite operator $G$. {Therefore, a $\pt$-symmetric operator $H$ with real eigenvalues is a quasi-Hermitian operator, and there exists a positive definite operator $\eta$ such that $H_\eta=\eta H\eta^{-1}$ is a Hermitian operator.}

$G$-pseudo-Hermitian operators can be shown to be self-adjoint with respect to a modified inner product defined with the operator $G$: $\miniprod{\cdot}{\cdot}_G:=\miniprod{\cdot}{G\cdot}$ as we show below
\begin{equation}
\begin{aligned}
\text{$\mathcal{O}^*$ is an adjoint of $\mathcal{O}$ if:}\hspace{0.5cm} \miniprod{\mathcal{O}^*\phi}{\psi}_G&= \miniprod{\phi}{\mathcal{O}\psi}_G\\
\Rightarrow\miniprod{\mathcal{O}^*\phi}{G\psi}&= \miniprod{\phi}{G\mathcal{O}\psi}\\
&=\miniprod{G^{-1}\mathcal{O}^\dagger G\phi}{G\psi}.
\end{aligned}
\end{equation}
Therefore, the adjoint of an operator $\mathcal{O}$ with respect to the inner product $\miniprod{\cdot}{\cdot}_G$ is given by $  \mathcal{O}^*=G^{-1}\mathcal{O}^\dagger G$ . The operator is self-adjoint if $\mathcal{O}=\mathcal{O}^*$, or equivalently $\mathcal{O}^\dagger G=G \mathcal{O}$, which is the $G$-pseudo-Hermiticity condition. When the operator $G$ is positive-definite, the inner product describes a physical Hilbert space. A Hilbert space where the inner product is given by $\miniprod{\cdot}{\cdot}_G$ will be denoted by $\h_G$, and the operator $G$ will be understood to be positive definite. The inner product with $G=\mathbb{1}$ will be referred to as the Euclidean inner product, and the corresponding Hilbert space $\h_{\mathbb{1}}$ will be denoted by $\h$ if the context is clear. 
\subsection{Hermitization of the operators} 
Given a state $\psi$ evolving under the G-pseudo-Hermitian Hamiltonian $H$, given by $\psi(t)=e^{-iHt}\psi(0)$, the norm of the state with respect to the inner product $\miniprod{\cdot}{\cdot}_G$ is time-invariant: $\partial_t\miniprod{\psi(t)}{\psi(t)}_G=0$. In this respect, this is a unitary evolution in $\h_G$. This also implies that there is a corresponding vector in Euclidean inner product space $\mathcal{H}$ whose norm is also time-invariant. Using the time invariance of the norm in $\mathcal{H}_G$ one can show
 \begin{equation}
 \begin{aligned}
%\partial_t  \miniprod{\psi(t)}{\psi(t)}_G=0 \\
%\Rightarrow \partial_t  \miniprod{\psi(t)}{G\psi(t)}= 0\\
%\Rightarrow 
\partial_t  \miniprod{\eta\psi(t)}{\eta\psi(t)}= 0.
\end{aligned}
 \end{equation}
Therefore, the vector $\psi_\eta(t):=\eta\psi(t)$ evolves unitarily in $\h$, under the Hermitian Hamiltonian $H_\eta=\eta H \eta^{-1}$.  As we have seen in the previous section, Hamiltonians with real eigenvalues can always be \textit{Hermitized}. Further note that the map $\h_G\to \h: \psi\to \eta\psi$, is a norm-preserving map between the two Hilbert spaces of equal dimension, i.e. $\miniprod{\psi_\eta}{\psi_\eta}=\miniprod{\psi}{\psi}_G$ and therefore, is a unitary map between the two Hilbert spaces. This also implies that the Hermitization map between the bounded linear operators $\mathcal{B}(\h_G)\to\mathcal{B}(\h)$ defined as follows
\begin{equation}\label{maptoeu}
\begin{aligned}
	\mathcal{O}\rightarrow \mathcal{O}_\eta & :=\eta\mathcal{O}\eta^{-1},
\end{aligned}
\end{equation}
is a norm-preserving transformation, which also preserves the self-adjointness: $\mathcal{O}^\dagger G= G \mathcal{O}\Leftrightarrow \mathcal{O}_\eta={\mathcal{O}_\eta}^\dagger$. The right and left eigenvectors, $\{\psi_i,\varphi_i\}$ respectively, of the quasi-Hermitian operator form the \textit{complete bi-orthonormal} basis for $\h_G$\ \cite{Brody16} with the following properties:
\begin{equation}\label{eq:biortho}
\begin{aligned}
	\miniprod{\psi_n}{\psi_m}_G=\miniprod{\varphi_n}{\psi_m}=&\delta_{mn}\, \text{\hspace{0.3cm} and\hspace{0.3cm}} \sum_n\ket{\psi_n}\bra{\varphi_n}=\mathbb{1}.
	\end{aligned}
\end{equation}
The quasi-Hermitian operator can be spectrally resolved as $\mathcal{O}=\sum_i p_i \ket{\psi_i}\bra{\varphi_i}$, and the metric operator can be constructed as $G=\sum_i \ket{\varphi_i}\bra{\varphi_i}$. Note that given the bi-orthonormal basis $\{\psi_i,\varphi_i\}$, the set $\{\eta\psi_i\}$ forms a complete orthonormal basis. Consequently, the spectral resolution of the pseudo-Hermitian operator is equivalent to the spectral resolution of the Hermitized operator $\mathcal{O}_\eta=\sum_n p_n \eta\ket{\psi_n}\bra{\psi_n}\eta$.  Finally, the map\ \eqref{maptoeu} also preserves the expectation value of an observable with respect to the transformed state. Given the density matrix $\bar{\rho}\in \h_G$ and the Hermitian density matrix $\rho_\eta=\eta\bar{\rho}\eta^{-1}$, we have
\begin{equation}
    \begin{aligned}
        \langle \mathcal{O}_\eta\rangle_{\rho_\eta}&=c_n\sum_n \miniprod{\eta\psi_n}{\mathcal{O}_\eta\eta\psi_n}\\
        &=c_n\sum_n \miniprod{\psi_n}{\eta^2\mathcal{O}\psi_n}=\langle \mathcal{O}\rangle_{\bar{\rho}}
    \end{aligned}
\end{equation}
We can also express this result as the following trace equality
\begin{equation}
\tr(\mathcal{O}_\eta\rho_\eta)=\tr(\mathcal{O}\bar{\rho}).
\end{equation}
This relation stems from the fact that the expectation value of an observable in any metric space can be given as a trace of the product of the observable with the density matrix, and that the trace operation gives a metric-independent answer. This is important enough to be stated as a distinct remark:
\\\\
\textbf{Remark:} \textit{the trace of any operator in Hilbert space $\mathcal{H}$ is equal to the trace of the same operator in $\mathcal{H}_G$}. 
\\\\
In summary, we can say that a system represented by the density matrix $\bar{\rho}$ on the Hilbert space $\h_G$ will have the same statistics as the system represented by the density matrix $\rho_\eta$ on $\h$ if the observables are Hermitized according to Eq.\eqref{maptoeu}.
\begin{figure}
    \centering
    \includegraphics[scale=0.15]{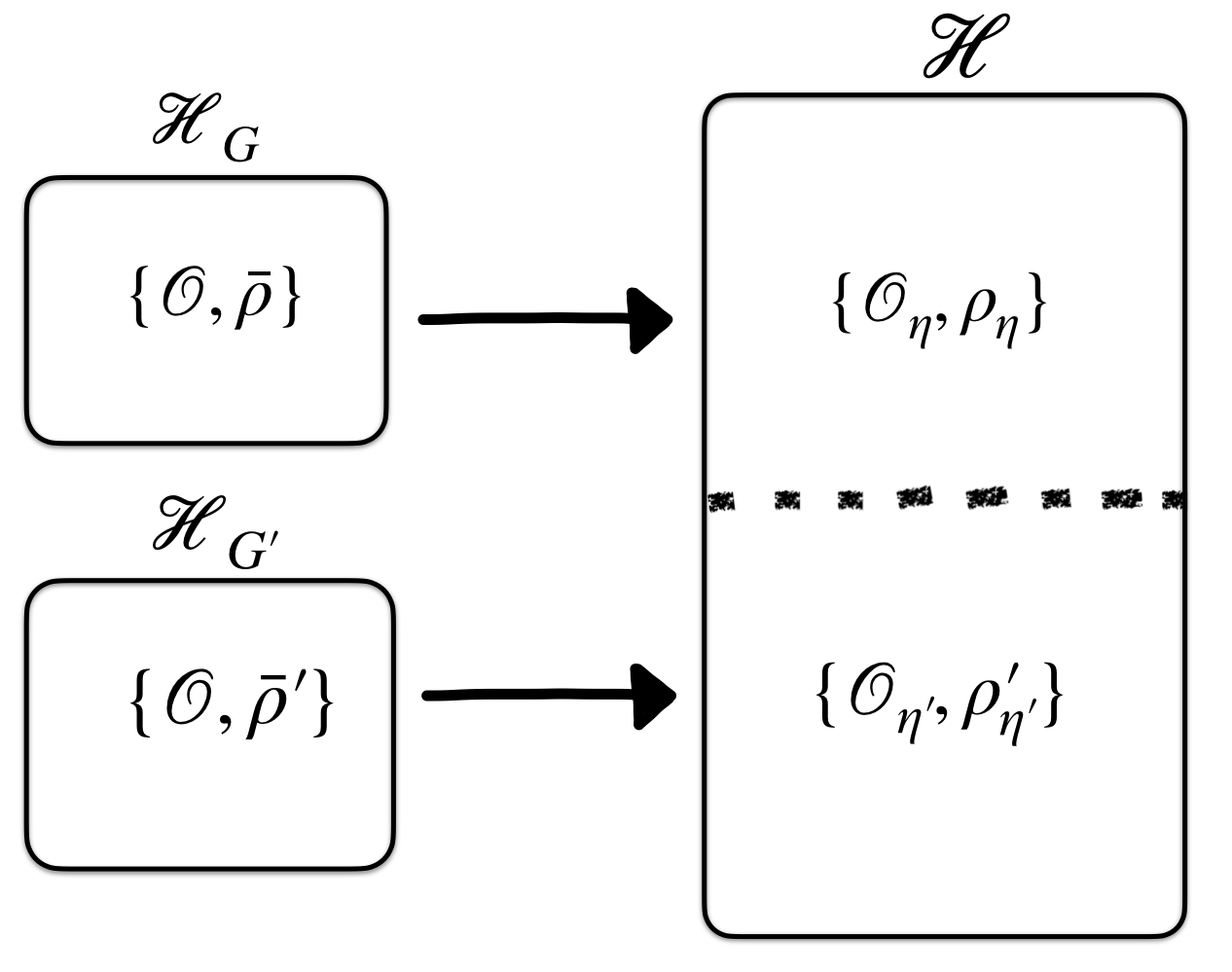}
    \caption{An unbroken $\pt$-symmetric evolution can be seen as unitary evolution under appropriate, non-unique choices of the metric, corresponding to evolutions in different Hilbert spaces $\mathcal{H}_G$ and $\mathcal{H}_{G'}$. Operators and states in these different Hilbert spaces can be mapped in the Euclidean ($G=\mathbb{1}$) Hilbert space to unitarily equivalent states and operators.}
    \label{fig:Hermitization}
\end{figure}

\subsection{Non-uniqueness of the metric and Hermitization}
Given a Hamiltonian $H$,  the compatible metric operator that satisfies the self-adjoint condition $H^\dagger G=GH$, is not unique. There is an infinite number of Hilbert spaces where the Hamiltonian is an observable for the energy of the system. However, all the choices of a compatible metric operator are equivalent and correspond to the same system. Given a Hamiltonian $H$ and the two compatible metric spaces $\h_G$ and $\h_{G'}$, the map $\mathcal{O}\to\mathcal{O}'=T^{-1}\mathcal{O}T$ is a self-adjointness preserving isomorphism between the bounded operators of the two spaces, where $T$ satisfies the relation $T^\dagger GT=G'$, or equivalently $[T,H]=0$. Therefore, both $\h_G$ and $\h_{G'}$ are valid physical Hilbert spaces for the evolution of a system under the Hamiltonian $H$. Hermitization of the operators from these two Hilbert spaces to the Euclidean metric Hilbert space $\h$, however, maps to a different state as we illustrate in fig. \ref{fig:Hermitization}. The operators $\mathcal{O}$ and $\mathcal{O}'$ will be mapped to the operators $\mathcal{O}_\eta=\eta\mathcal{O}\eta^{-1}$ and $\mathcal{O}'_{\eta'}=\eta'\mathcal{O}'\eta'^{-1}$ respectively, where $\eta=\sqrt{G}$ and $\eta'=\sqrt{G'}$. Since both the change of the metric space and the Hermitization do not change the statistics of the system, one can conclude that the expectation value of the observable, when Hermitized from different metric spaces, remains the same, as we show in the following.
\begin{equation}
\begin{aligned}
	\tr(\mathcal{O}_\eta\rho_\eta)&=\tr(\mathcal{O}\bar{\rho})\\
 &=\tr(\mathcal{O}'\bar{\rho}')\\
 &=\tr(\eta' \mathcal{O}'\eta'^{-1}\eta'\rho'\eta'^{-1})
=\tr(\mathcal{O}'_{\eta'}\rho_{\eta'}').
\end{aligned}
\end{equation}
As we learned that the set $\{\eta\psi_n\}$ forms a complete orthonormal basis of $\h$, the set $\{\eta'T^{-1}\psi_n\}$ will form another complete orthonormal basis of $\h$. Surely, any two complete orthonormal sets of basis vectors are related by a unitary, hence there exists a unitary operator  $U$ on $\h$ such that $\eta'T^{-1}=U\eta$. From here, we can conclude that the operators $\mathcal{O}_\eta$ and $\mathcal{O}'_{\eta'}$ on $\h$ are also unitarily equivalent
\begin{equation}\label{unitaryeqv}
    \begin{aligned}
            \mathcal{O}'_{\eta'}= U \mathcal{O}_{\eta}U^\dagger  .     
    \end{aligned}
\end{equation}
\section{Subsystems in $\h_G$}\label{sec:subsystems}
In this subsection, we will  propose that the partial trace of the Hermitized system $\rho_\eta$ in $\h=\h_A\otimes\h_B$ is the correct way to describe the subsystems $A$ and $B$ of $\bar{\rho}$ in $\h_{G_{AB}}$. The partial trace in $\h_{G_{AB}}$ is well defined if the metric if of the form $G_{AB}=G_A\otimes G_B$, which results in a tensor product Hilbert space structure $\h_{G_A}\otimes\h_{G_B}$. The partial trace over the system $A$ gives a reduced density matrix in $\h_{G_B}$ as follows:
\begin{equation}
\begin{aligned}
\bar{\rho}_B&=\sum_i (\bra{\psi_n}\otimes I_B)(\bar{\rho}_{AB})(\ket{\psi_n}_{G_A}\otimes I_B)
\end{aligned}
\end{equation}
The state $\bar{\rho}_B$ $\in$ $\mathcal{H}_{G_{B}}$ is mapped to the Euclidean inner product space $\mathcal{H}_{B}$ as $\rho_B^\eta:=\eta_B\bar{\rho}_B\eta_B^{-1}$. One can show that the reduced state $\rho_B^\eta$ can also be constructed by first Hermitizing the state $\bar{\rho}_{AB}$ and then performing the partial trace in $\h_{AB}$, and therefore the two methods of partial trace are equivalent:
\begin{equation}\label{reduced}
\begin{aligned}    
\rho_B^\eta &=\tr_A((\eta_{A}\otimes \eta_{B}) \bar{\rho}_{AB}(\eta_{A}\otimes \eta_{B})^{-1})\\
&=\eta_B \tr_A((\eta_{A}\otimes I_{B})\ \bar{\rho}_{AB}\ \eta_{A}^{-1}\otimes I_B)\eta_{B}^{-1}\\
&=\eta_B\bar{\rho}_B\ \eta_B^{-1}
\end{aligned}
\end{equation}
Under a different choice of the metric, as long as the new metric is still in the tensor product form, the reduced state $\bar{\rho}_B$ is uniquely defined and its map to the Euclidean Hilbert space is unitarily equivalent to $\rho_B^\eta$. The reduced dynamical map, which evolves the state $\bar{\rho}_B(t)$ in time uniquely defined in $\mathcal{H}_{G_B}$ and is defined in $\mathcal{H}_B$ up to unitary equivalence in $\h_B$. 

The equivalence between the partial trace after Hermitization and without Hermitization for such metric spaces can also be seen from an operational point of view. Consider a composite system with position and spin degrees of freedom. Let $\hat{X}$ and $\sigma_3$ be the Hermitian operators with eigenstates $\ket{x}$ and $\ket{s}$ with eigenvalues $x$ and $s$, respectively. Let the set $\{\ket{x,s}\}$ form the  complete orthonormal basis of the Euclidean metric space $\h_p\otimes\h_s$. We define  $\ket{\psi(x,s)}=\eta^{-1}\ket{x,s}$, so that the set $\{\psi(x,s),\eta^2\psi(x,s)\}$ forms a complete bi-orthonormal basis in $\mathcal{H}_G$ and are also the eigenvectors of the corresponding quasi-Hermitian position and spin observables in $\mathcal{H}_G$ given by $\overline{\hat{X}}=\eta^{-1}(\hat{X}\otimes I_s)\eta$ and $\overline{\sigma}_3=\eta^{-1}(I_p\otimes \sigma_3)\eta$  respectively. Now, given a density matrix $\rho_\eta\in\h$, the terms $\bra{x,s}\rho_\eta\ket{x',s'}$ are the matrix elements with respect to the given basis. One can show that they are exactly equal to the matrix elements of the corresponding density matrix $\bar{\rho}\in\h_G$ with respect to the bi-orthonormal basis of $\h_G$: 
\begin{equation}
\begin{aligned}
    \bar{\rho}(x,x';s,s')&:=\bra{\psi(x,s)}\bar{\rho}\ket{\psi(x',s')}_G\\
    &= \rho_\eta(x,x';s,s'):=\bra{x,s}\rho_\eta\ket{x',s'}.
\end{aligned}
\end{equation}
Note that the probability of getting the spin value $s$, while measuring in the $\sigma_3$ direction, at the lattice point $x$ is given by the element 
\begin{equation}
\begin{aligned}
    p(s,x|n_3)=\bar{\rho}(x,x;s,s)
    &=\rho_\eta(x,x;s.s).
\end{aligned}
\end{equation}
Spin expectation values in other directions can be given by an appropriate unitary transformation of $\rho_\eta$. For instance, the probability of getting spin $s$ at a lattice point $x$ in the $\hat{n}_1$ direction (corresponding to the measurement of $I_N\otimes\sigma_1$ in $\h$ or the measurement of $\bar{\sigma}_1$ in $\h_G$) is given by 
\begin{equation}
    \begin{aligned}
p(s,x|\hat{n}_1)=\bra{\psi(x,s)}\overline{U}_1^*\ \bar{\rho}\ \overline{U}_1 \ket{\psi(x,s)}_G= \bra{x,s}U_1^\dagger \rho_\eta U_1)\ket{x,s},
    \end{aligned}
\end{equation}
where $U_1$ is a unitary on $\h$ and $\overline{U}_1=\eta^{-1} U \eta$ is a unitary on $\h_G$ and $\overline{U}^*=G^{-1}U^\dagger G$ is the adjoint operation in $\h_G$. The statistics of the spin degrees of freedom, obtained by summing over the position degrees of freedom, can therefore be either from the state $\bar{\rho}$ or its Hermitized version $\rho_\eta$:
\begin{equation}
\begin{aligned}
    p(s|\hat{n}_i)&=\sum_x p(s,x|\hat{n}_i)\\
    &=\sum_x\bra{\psi(x,s)}\overline{U}_i^*\ \bar{\rho}\ \overline{U}_i \ket{\psi(x,s)}_G.
\end{aligned}
  \end{equation}
However, these statistics are faithfully represented by the state $\rho_\eta^c=\tr_A(\rho_\eta)$ as all the probabilities $p(s|\hat{n}_i)$ can be extracted from $\rho_\eta^c$ as follows
\begin{equation}
    p(s|\hat{n}_i)=\bra{s}{U_i^c}^\dagger\rho_\eta^c\ {U_i^c} \ket{s},
\end{equation}
where, $U_i^c$ are the appropriate unitary operators such that ${U_i^c} \ket{s}$ is the eigenvector of $\sigma_i$. For example $U_3^c=I_2$ and $U_1^c=\dfrac{(\mathbb{1}-i\sigma_2)}{\sqrt{2}}$. This discussion indicates that partial tracing after Hermitization of the state represented by the density matrix $\bar{\rho}\in\h_G$ is a well-defined way to describe the subsystem 

\subsection{Metric dependence of the subsystem:} As we have discussed, the freedom of choice of the Hamiltonian compatible metric does not change the properties of the system in terms of the expectation values of the observables. However, this choice does affect the subsystem's properties.  Let $\bar{\rho}\in\mathcal{H}_G$ be the density matrix, we know that $\overline{\rho}'=T^{-1}\overline{\rho}T$ is the corresponding state in $\mathcal{H}_{G'}$, where $T^\dagger GT=G'$ and $[T,H]=0$. Hermitizing these states, we get $\rho_\eta$ and ${\rho}_{\eta'}$ respectively, that are related by $\rho_{\eta'}=U\rho_\eta U^\dagger$ (from Eq. \eqref{unitaryeqv}). Here, the unitary $U$ depends on the metric operators $G$ and $G'$ as $U=\eta'T^{-1}\eta^{-1}$. 
    
\begin{theorem}
Given the states $\bar{\rho}\in\h_G$ and $\bar{\rho}'=T^{-1}\bar{\rho}T\in\h_{G'}$, where $T^\dagger G T=G'$, the properties of the subsystems given by $\rho_\eta^c=\tr_p(\rho_\eta)$ and $\rho_{\eta'}^c=\tr_p(\rho_{\eta'}')$ are metric dependent unless both $\h_G$ and $\h_{G'}$ have a tensor product decomposition.
\end{theorem}
\begin{proof}
If the $\h_G$ and $\h_{G'}$ both have a tensor product decomposition of the form $\h_{G_p}\otimes\h_{G_c}$ (which is possible only if the metric operators have a tensor product form), then $U$ can be a local unitary of the form $U_p\otimes U_c$ . In such cases, the reduced states are also related by a unitary, ${\rho}_{\eta'}^c=U_c\rho_\eta^c U_c^\dagger$and the same is true for all the observables on $\h_G$ and $\h_{G'}$. Therefore, the statistics of the states are independent of the choice of the metric
\begin{equation}
    \begin{aligned}
  \tr(\mathcal{O}_{\eta'}\rho_{\eta'}^c)&=\tr(U_c\mathcal{O}_{\eta}U_c^\dagger U_c\rho_{\eta}^cU_c)\\
  &=\tr(\mathcal{O}_\eta\rho_\eta^c)
    \end{aligned}
    \end{equation}
However, if the metric operators do not have a tensor product form, the reduced states $\rho^c_{\eta_c}$ and $\rho^c_{\eta'}$  are not unitarily equivalent, and the above Eq. does not hold. In that case, the statistics of the reduced state depend on the choice of the Hamiltonian compatible metric operator.
\end{proof}
\noindent
\textit{Observable dependence}: From this perspective, the choice of the metric is equivalent to the choice of a frame of reference for the system. It is well understood that while a global unitary transformation on the system does not change the expectation values of the observables on the system, such a transformation does affect the entanglement structure of the system. The above theorem might lead one to conclude that there exists a preferred choice of the metric for which the reduced dynamics is well defined in both $\mathcal{H}_G$ and $\mathcal{H}$, which is the metric in tensor product form. However, this would be a misguided conclusion since the choice of a particular form of the inner product is not guided by any physical principle. As noted in\ \cite{scholtz}, the metric operator is fixed by a choice of a complete set of irreducible observables. Given a set of observables $\{\mathcal{O}_i\}$  on $\h_G$ and $\h_G'$ they must satisfy $[\mathcal{O}_i,T]$ where $T^\dagger GT=G'$. Therefore, the metric space where a set of operators $\{\mathcal{O}_i\}$, a well-defined set of observables, is uniquely fixed if $T=\mathbb{1}$, that is, the set $\{\mathcal{O}_i\}$ is irreducible. This tells us that the decomposition of a subsystem can be observable dependent, a notion that has been studied in its generality before, e.g.\ \cite{paolo01,paolo03,Ahmed22}. In summary, the metric dependence of the subsystem is very much in line with the usual quantum mechanics. At the same time, it offers a unique window into the physics of $\pt$-symmetric quantum mechanics. 
%===================================================================
\section{Subsystems of a quantum walk and the effect of the metric}\label{sec:qw}
%====================================================================
Most $\pt$-symmetric non-Hermitian Hamiltonians would typically not be compatible with a metric in tensor product form. Here, we will study the evolution of a composite particle under such a Hamiltonian. Unitary evolution of a quantum particle on a lattice can be described by a quantum walk\ \citep{meyer96} which can be seen as a composite system, with its position and spin degrees of freedom as its subsystems described in the Euclidean Hilbert space $\h=\h_p\otimes \h_s$. These are being increasingly used as a tool for quantum simulation\ \cite{Childs09}. With the periodic boundary condition on the lattice, we choose a walk that diagonalizes in the Fourier basis, and the Shift and Coin operators can be written as
\begin{equation}
    \begin{aligned}
	S=&\sum_{k}\ket{k}\bra{k}\otimes \begin{pmatrix}
e^{ik} & 0\\
0 & e^{-ik}
\end{pmatrix}= \oplus_{k}S(k)\\
\overline{C}(\theta)&=\mathbb{1}_N\otimes\begin{pmatrix}
\cos{\theta} & i\sin{\theta}\\
i\sin{\theta} & \cos{\theta}
\end{pmatrix}=\oplus_k C(\theta)
\end{aligned}
\end{equation}
The $\pt$-symmetric non-unitary quantum walks have been defined in\ \cite{Ken16} based on a balanced gain-loss model, which in the momentum space remains diagonal:
\begin{equation}\label{eq:nhqwalk}
\begin{aligned}	
& W_{nh}=\oplus_kW_c(k),\\
\text{where}\hspace{0.3cm}
&W_c(k)=C(\theta_1/2)S(k) \G^{-1}C(\theta_2)S(k)\G C(\theta_1/2),
\end{aligned}
\end{equation}
where the non-unitary contribution from the operators $\G$ simulating gain and loss is given by
\begin{equation}
    \begin{aligned}
       \G&=\begin{pmatrix}
\delta & 0\\
0 & 1/\delta
\end{pmatrix},\hspace{1cm} \delta\in\mathbb{R}^+,
    \end{aligned}
\end{equation}
The $\pt$-symmetric nature of this evolution can be deduced from the fact that $W_c(k)$ in Eq.\ \eqref{eq:nhqwalk} is inverted under the charge conjugation, which is an antilinear operation. In the appendix\ \ref{app:QWnon-sep}, we construct the most general metric operator for the above quantum walk and show the following
\\[0.2cm]
\textbf{Claim:} \textit{Given the $\pt$-symmetric Hamiltonian for the quantum walk operator in Eq. \eqref{eq:nhqwalk}, if $\delta\neq 1$, then there does not exist a metric operator which is separable in the internal and external degrees of freedom.}
\\[0.2cm]
As a result, the dynamics of a coin state, with position/momentum degrees of freedom traced out, is not well defined on the Hilbert space $\mathcal{H}_G$ for any $G$ compatible with the walk operator. Therefore, the subsystem dynamics in the Euclidean Hilbert space $\mathcal{H}$ depend on the choice of the metric operator.
\\\\
In this section, we will study two dynamical features of the coin state: the non-Markovianity of the reduced dynamics and the entanglement growth between the coin and position degrees of freedom (by calculating the von Neumann entropy of the reduced coin state). The evolution of a subsystem can rarely be defined as a CPTP evolution, the main reason being that there is an information backflow from one system to the other, resulting in a non-Markovian evolution that cannot be modeled by a CPTP map. The characterization of the non-Markovianity of the subsystem dynamics, therefore, is an important aspect of the subsystem dynamics. Similarly, the entropy of the reduced states, which is also an entanglement measure for pure bipartite states, is another important characteristic of the subsystem that can also affect the non-Markovianity of the dynamics. We study these properties of the reduced dynamics under different metric operators. We will restrict ourselves to a metric that also preserves the direct sum decomposition of the Hilbert space and, therefore, takes the form $G=\oplus_k G_c(k,x,y)$ where metric operators $G_c(k,x,y)$ on $\mathbb{C}^2$ are given by
\begin{equation}
    G_c(k,x,y)=x(\ket{r_+(k)}\bra{r_+(k)}+y\ket{r_-(k)}\bra{r_-(k)}),
\end{equation}
Here $\ket{r_{\pm}(k)}$ are the left eigenvectors of the $\pt$-symmetric walk operator and $x,y\in \mathbb{R}^+$. In this work, we choose three different metric operators of the direct sum form: first with $x(k)=y(k)=1\forall k$, denoted by $G_1$, and the other two with $x(k)$ and $y(k)$ chosen randomly from a uniform probability distribution denoted by $G_2$ and $G_3$.  Further, we will study the walk at three different degrees of non-Hermiticity: $\delta=1$ (Hermitian), $\delta=1.2$, and $\delta=1.3$. Note that with the $\theta_1$ and $\theta_2$ fixed to $\pi/4$ and $-\pi/7$, the exceptional point lies at $\delta_{PT}\sim 1.35$.
\subsubsection{Information backflow to the coin state} 
\noindent
We will study the information backflow during the dynamics of the coin state of the Hermitized quantum walk. The reduced dynamical map for the Hermitized state is given by
\begin{equation}	
\Lambda_\eta(\rho_c):=\rho_c(t)=\tr_p(W_\eta^t\rho_\eta {W_\eta^\dagger}^t), 
\end{equation}
where $W_\eta=\eta W_{nh}\eta^{-1}$ for the $\pt$-symmetric walk $W_{nh}$ defined in Eq. 
\eqref{eq:nhqwalk}. Given a state $\bar{\rho}\in \mathcal{H}_G$, we evolve a corresponding state $\rho_\eta=\eta\bar{\rho}\eta^{-1}$ under this operator. The increase in the trace distance between two states under a map indicates a backflow of information to the system\ \cite{blprev}, since the trace distance, given by $D(\rho,\sigma)=\dfrac{1}{2}\sum_i|\lambda_i|$, where $\lambda_i$ are the eigenvalues of the operator $\rho-\sigma$, is a measure of distinguishability between the states and should always decrease under the loss of information. Summing over the increase in distinguishability for a particular initial set of states, and then maximizing over the initial states, gives us the Breuer-Lane-Piilo (BLP) measure.

\begin{center}
    \begin{figure}[!ht]
    \centering
    \includegraphics[scale=0.25]{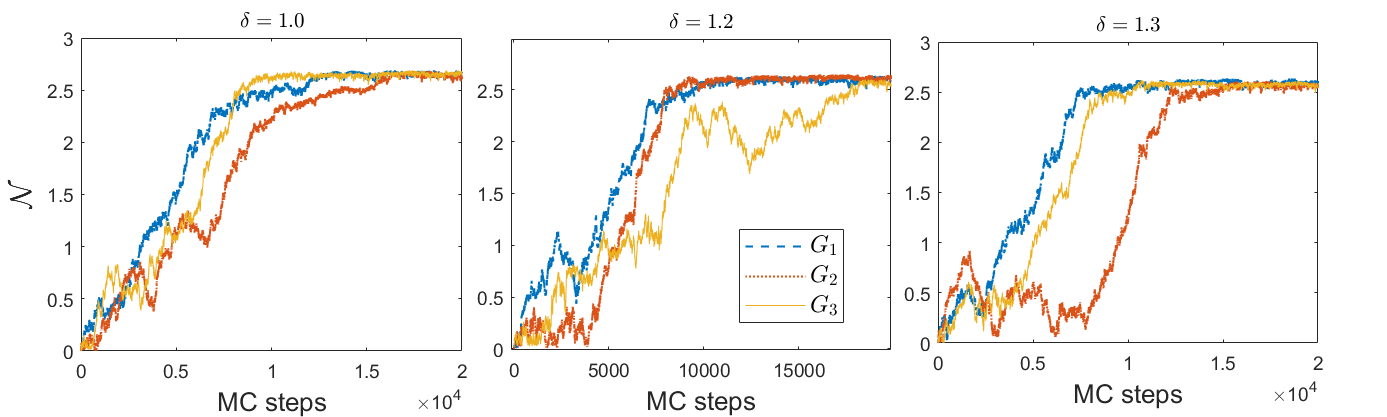}
    \caption{Maximization of the BLP measure $N$ using the simulated annealing method.  Each plot marks the measure evaluated after 50 steps of the walk for 3 different metric operators $G_1$, $G_2$, and $G_3$. We see that the measures saturate at roughly the same values. The Monte-Carlo method has finite precision, and within this error, the information backflow to the coin state is independent of the metric. The effect of the chosen metric is visible in another measure of non-Markovianity.}\label{blpfig}
\end{figure}
\end{center}

Let $D(\Lambda(\rho),\Lambda(\sigma))-D(\rho,\sigma)=\bigtriangledown$ denote the change in the trace distance due to the discrete map $\Lambda$ given either under the state normalization method or the metric formalism.
The measure of information back-flow, depending on the initial states $\rho$ and $\sigma$, for the discrete dynamics, is given by:
\begin{equation}
\begin{aligned}
N(t,\rho,\sigma)=
\begin{cases}
& N(t-1,\rho,\sigma)+\bigtriangledown, \hspace{.5cm} \text{if $\bigtriangledown>0$}\\
& N(t-1,\rho,\sigma), \hspace{.5cm} \text{if $\bigtriangledown\le 0$}
\end{cases}
\end{aligned}
\end{equation}
where $N(0,\rho,\sigma)=0$. If one maximizes over the set of all initial states $\rho$ and $\sigma$, we get a measure of information back-flow which is a characteristic of the map $\Lambda$ alone:
\begin{equation}
	\mathcal{N}_t:=\max_{\rho,\sigma}\{N(t,\rho,\sigma)\}.
\end{equation}
\textit{Case of tensor product metric operator: }If the metric operator is in a tensor product form such that $\sqrt{G}=\eta=\eta_p\otimes\eta_c$, then we know that the reduced dynamics in $\mathcal{H}_{G_c}$, given by $\Lambda=\tr_p(W^t\bar{\rho}{W^\#}^t)$, is well defined. For the state $\rho_{\eta_c}:=\eta_c\bar{\rho}_c\eta_c^{-1}$, one can show that $\Lambda_{\eta}(\rho_{\eta_c})=\eta_c \Lambda(\bar{\rho}_c) \eta_c^{-1}$ in the following way
\begin{equation}
    \begin{aligned}
        \Lambda_\eta(\rho_c)&=\tr_p(W_\eta^t\rho_\eta {W_\eta^\dagger}^t) \\
        &=\tr_p(\eta W^t\bar{\rho} \eta^{-2}{W^\dagger}^t \eta) \\
        &=\tr_p(\eta_p\otimes\eta_c W^t\bar{\rho} {W^\#}^t\eta_p^{-1}\otimes\eta_c^{-1} ) \\
         &=\eta_c \tr_p(W^t\bar{\rho} {W^\#}^t)\eta_c^{-1} =\eta_c \Lambda(\bar{\rho}_c)\eta_c^{-1}
    \end{aligned}
\end{equation}
This also implies that the states $\Lambda_{\eta}(\rho_{\eta_c})$ and $\Lambda_{\eta'}(\rho'_{\eta_c'})$ are unitarily equivalent:  
\begin{equation}\label{reducedunitaryeqv}
    \Lambda_{\eta'}(\rho'_{\eta_c'})=U\Lambda_{\eta}(\rho_{\eta_c}) U^\dagger.
\end{equation} 
Since trace distance is invariant under unitary transformations, we have the following identity:
\begin{equation}
    D(\Lambda_{\eta}(\rho_{\eta_c}),\Lambda_{\eta}(\sigma_{\eta_c}))=D(\Lambda_{\eta'}(\rho'_{\eta'_c}),\Lambda_{\eta'}(\sigma'_{\eta'_c})).
\end{equation}
Therefore, for metric operators with tensor product decomposition, we have the following results: 
\begin{enumerate}
    \item The BLP measure is independent of the metric chosen
    \item If the pair of states $(\rho_{\eta_c},\sigma_{\eta_c})$ maximizes the measure, the pairs of states $(\rho'_{\eta'_c},\sigma'_{\eta'_c})$ will also maximize the measure.
\end{enumerate}
For the metric operator not in the tensor product form, we do not have states like $\rho_{\eta_c}$, and therefore we will denote by $\rho_\eta^c$ the states that are constructed by a partial trace of states $\rho_\eta\in\mathcal{H}$. We maximize the measure over the set of states by using the simulated annealing technique and calculate the measure for three different metric operators. The results are presented in Figure\ \ref{blpfig}.
\begin{enumerate}
    \item The BLP measure, within the error margin of $10^{-2}$ units, does not change with the metric or even with the non-Hermiticity parameter $\delta$. 
    \item The measure for a given $\delta$ evolution {can be} maximized for unitarily inequivalent states (within the error of $10^{-2}$). The table below shows the trace distance between the  states for which the maximum BLP measure is achieved, $(\rho_{\eta_1},\sigma_{\eta_1})$ and $(\rho'_{\eta'},\sigma'_{\eta'})$ where $\eta'$ is the positive square root of $G'\in\{G_2,G_3\}$. {Note that if $\rho_\eta$ and $\rho'_{\eta'}$ are unitarily equivalent, we must have $D(\rho_\eta,\rho'_{\eta'})=0$, or more precisely $D(\rho_\eta,\rho'_{\eta'})\lessapprox 0.02$, which is the minimum error in the maximization algorithm. The table~\ref{tab:my_label} shows that the states which maximize the BLP measure for three different metric operators are unitarily inequivalent.} 
\\[2mm]
\begin{table}Trace distance between optimal states
    \centering
    \begin{tabular}{ |p{0.7cm}||p{1.8cm}|p{1.8cm}||p{1.8cm}|p{1.8cm}|  }
 \hline
 ~$\delta$& $D(\rho_{\eta_1},\rho_{\eta_2})$  &$D(\sigma_{\eta_1},\sigma_{\eta_2})$& $D(\rho_{\eta_1},\rho_{\eta_3})$  &$D(\sigma_{\eta_1},\sigma_{\eta_3})$\\
 \hline
 1.0&0.0340&0.0024&0.0170&0.0170\\
 1.2&0.0180&0.0205&0.0380&0.0399\\
 1.3&0.0759&0.0299&0.0741&0.1080\\
 \hline
\end{tabular}
    \caption{{The table shows the distance between the pairs that maximize the BLP measure for different metric operators and non-Hermiticity parameters $\delta$. Here, the pair $(\rho_{\eta_i},\sigma_{\eta_i})$ maximizes the BLP measure for the metric operator $G_i$, where $i\in\{1,2,3\}$}}
    \label{tab:my_label}
\end{table}
\end{enumerate}
The above two results suggest that while the change in the metric might not have any effect on the information backflow to the subsystem, the states for which the measure is maximized do depend on the choice of the metric.
\subsubsection{CP- indivisibility}
In this section, we will study the non-Markovianity of a process through the violation of the divisibility of the quantum map. A quantum Markovian process between two times $t_0$ to $t$, represented by a completely positive trace preserving (CPTP) map $\mathcal{E}_{t,t_0}$, can be broken down into the composition of two maps:
\begin{equation}\label{mapcomp}
	\mathcal{E}_{t, t_0} = \mathcal{E}_{t, t_1}\circ\mathcal{E}_{t_1, t_0},
\end{equation}
for all times $t\le t_1 \le t_0$. If the intermediate map $\mathcal{E}_{t,t_1}$ is a completely positive (CP) map for all times, the process is called a CP-divisible process. Based on the fact that the local CPTP maps do not increase entanglement between two systems, one can detect the lack of complete positivity of the intermediate map $\mathcal{E}_{t,t_1}$. The Choi matrix of  a completely positive map has a trace norm $1$, otherwise, it is greater than $1$. Using this property of the CP maps, the Rivas-Huelga-Plenio (RHP) measure\, \cite{rhp} of the dynamical map $\Lambda$ is given by:
\begin{equation}\label{rhpdef}
\begin{aligned}	                                                    \mathcal{I}_{RHP}&=\sum_t g(t),\\
\text{where}\hspace{1cm} g(t)&=||\mathbb{1}\otimes\Lambda_\eta(t+1,t)\ket{\Phi}\bra{\Phi}\ ||_1-1
\end{aligned}
\end{equation}
and $\ket{\Phi}=\sum_i \ket{i}\ket{i}$ is the maximally entangled vector in $\h$ and $\Lambda_\eta(t+1,t)$ is the intermediate map from time $t$ to $t+1$ on the coin state. The operator $C_{\Lambda_\eta}=\mathbb{1}\otimes\Lambda_\eta(t+1,t)\ket{\Phi}\bra{\Phi}$ is called the Choi matrix of the map $\Lambda_\eta(t+1,t)$. We outline its construction in the appendix\ \ref{app:intermap}. If the intermediate map is completely positive, then $g(t)=0$. 

\begin{figure}[!h]
\begin{subfigure}{.5\textwidth}
  \centering
   \includegraphics[scale=0.25]{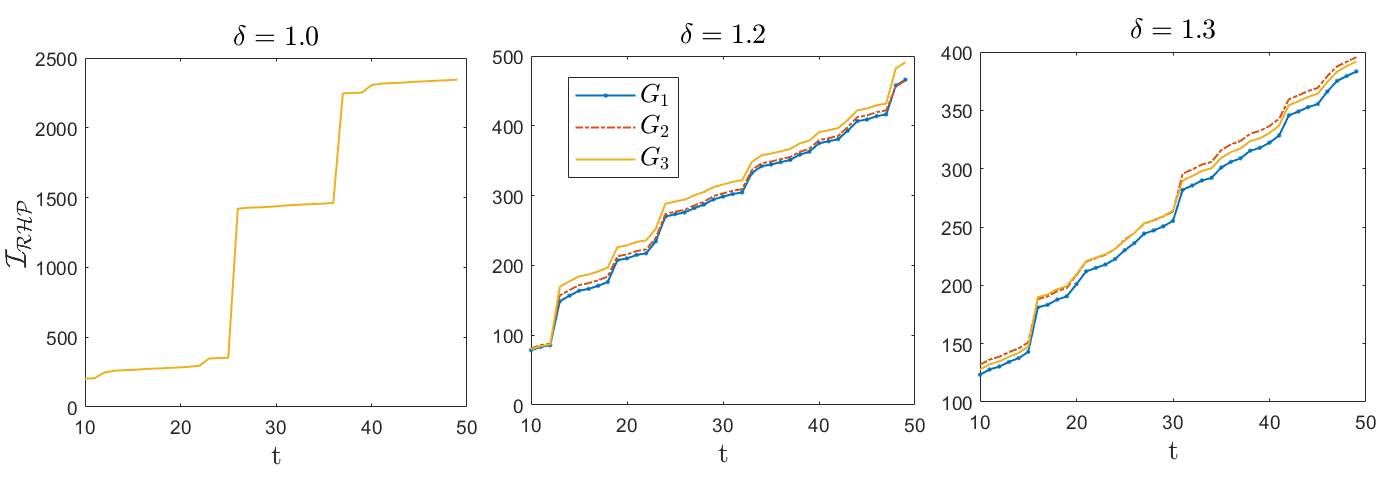}
       \caption{} \label{rhpfig}
\end{subfigure}
\begin{subfigure}{.5\textwidth}
  \centering
  \includegraphics[scale=0.25]{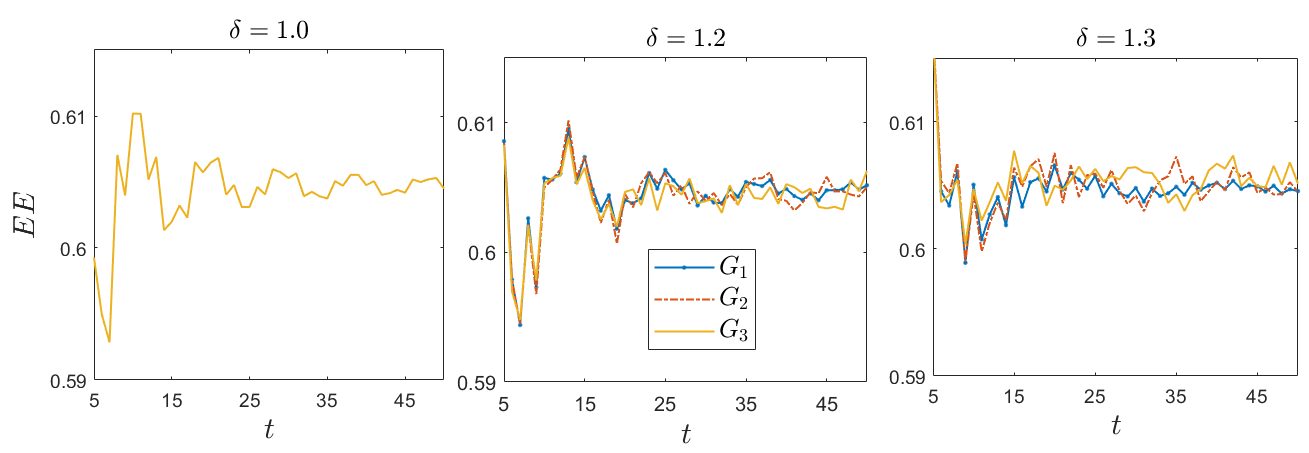}
    \caption{} \label{entfig}
\end{subfigure}
 \caption{\ref{rhpfig} RHP measure as a function of time for the reduced dynamics under different choices of metric operators. We see clearly that the CP-indivisibility of the coin state's dynamical map in the Euclidean metric space has different measure values for different choices of the metric in the $\delta\neq 1$ case. This is a reflection of the non-separability of the metric operator in the non-trivial metric space. For the Hermitian case, we do not see any metric dependency. \ref{entfig} Coin-position entanglement as a function of time under different choices of the metric operator. The coin-position entanglement in the Euclidean metric space shows the metric dependency for dynamics with $\delta\neq 1$.}
\end{figure}

\textit{Case of tensor product metric operator:} If the metric $G=\eta^2$ and $G'=\eta'^2$ are both in tensor product form, then one can show that the RHP measure would be independent of the metric chosen. This is because the maximally entangled states are invariant under local unitaries $U_1\otimes U_2\ket{\Phi}=\ket{\Phi'}$. Therefore, the Choi matrices of the Hermitized reduced maps are unitarily equivalent, as we show below:
\begin{equation}
\begin{aligned}
     &\mathbb{1}\otimes\Lambda_{\eta'}(t+1,t)\big(\ket{\Phi'}\bra{\Phi'}\big)\\
     &= \mathbb{1}\otimes\Lambda_{\eta'}(t+1,t)\big(U_1\otimes U_2 \ket{\Phi}\bra{\Phi}U_1^\dagger\otimes U_2^\dagger\big)\\
     &=\sum_{i,j} U_1\ket{i}\bra{j}U_1^\dagger \otimes \Lambda_{\eta'}(t+1,t)(U_2\ket{i}\bra{j}U_2^\dagger)\\
     &=\sum_{i,j} U_1\ket{i}\bra{j}U_1^\dagger \otimes U_2\Lambda_{\eta}(t+1,t)(\ket{i}\bra{j})U_2^\dagger\\
    &= (U_1\otimes U_2) \big(\mathbb{1}\otimes\Lambda_{\eta}(t+1,t)\big(\ket{\Phi}\bra{\Phi}\big)\big) (U_1^\dagger \otimes U_2^\dagger).
\end{aligned}
\end{equation}
In the third equality, we have used the result from Eq. \eqref{reducedunitaryeqv}. Since the trace norm is invariant under unitary transformation, the function $g(t)$ and, consequently, the RHP measure is invariant under the metric choice as long as the chosen metric operators are in tensor product form. This, however, is not the case for the metric operators not in the tensor product form, as we show through the example of the quantum walk.
\\\\
Figure \ref{rhpfig} shows the increase in the RHP measure with time for different $\delta$ parameters and for different metric choices. We note that, for $\delta=1$, that is, in the Hermitian case, the measure is independent of the metric. While the Hermitian Hamiltonian case does allow for the tensor product structure of the metric, we have chosen metric operators that are not in this separable form (specifically, $G_2$ and $G_3$). For the other two cases of $\delta\neq 0$, the measure shows variations with the choice of the metric. This is not surprising since RHP is an entanglement-based measure and the entanglement structure of the state is metric dependent as shown in Figure\ \ref{entfig}.
Since the evolution is unitary, the von Neumann entropy of the reduced state captures the entanglement between the position and coin space. Other works in the past have studied other measures of bipartite entanglement under a non-Hermitian Hamiltonian, interpreting the evolution to that of an open system \cite{De23}. Note that for the case of tensor-product metric operators $G_1\otimes G_2$, the entanglement entropy of the states  $\rho_\eta$ and $\rho'_{\eta'}$ is metric invariant, since the reduced states under two different metric operators are unitarily equivalent and therefore will have the same von Neumann entropy. In the case of metric operators not of the form $G_1\otimes G_2$, the entanglement entropy changes with the choice of the metric as shown in Figure \ref{entfig}. As was discussed in the earlier sections, this is expected, since the states $\rho_{\eta_1}$ and $\rho'_{\eta_2}$ are related by a global unitary $\rho_{\eta_1}=U\rho_{\eta_2}U^\dagger$, and therefore correspond to different bipartitions.
%============== ===================
%============== ===================
\section{Conclusions}\label{sec6}
In this work, we have investigated the problem of defining the reduced dynamics of unbroken $\pt$-symmetric dynamics in the pseudo-Hermitian quantum mechanics framework. When the metric operator $G$, compatible with the Hamiltonian, does not allow a particular tensor product decomposition ($G\neq G_A\otimes G_B$), the partial trace operation is not well defined in the metric space $\h_G$. We have argued that the Hermitization of the states and observables, followed by partial trace, is a consistent method of defining subsystems for such systems. This immediately also leads us to conclude that the properties of such subsystems should depend on the choice of the Hamiltonian-compatible metric operator, especially when the chosen metric operators are not in the tensor product form. While the freedom in the choice of the metric is of no consequence when one deals with the system itself, the subsystem decomposition depends on this choice.  The choice of the metric also stems from a particular choice of the observables, and it is understood that the partitioning of a system is observable-induced.  This is equivalent to the common observation that the choice of a different frame of reference for the system, while not affecting the system's statistics, does affect the subsystem decomposition. This becomes clear as we study the CP indivisibility of the reduced dynamics and the changes in the von Neumann entropy of the reduced state under the reduced dynamics of a quantum walk. These properties depend on the choice of the metric in non-tensor-product form, indicating that such choices lead to a change in the partitioning of the system.
\\
The metric dependency of non-Markovianity, bipartite entanglement, and other subsystem properties can be proven to be useful when there is a need to manipulate these properties. The operational interpretation of the metric operator means that an experimentalist can have control over the dynamics of the subsystems simply by changing the set of observables.
%===================================
\noindent
\\
\section*{Acknowledgment}
We thank Dr. Sibasish Ghosh for stimulating conversation and clarifications on metric formalism. 
%==============================================

\appendix
\section{Non-separability of the quantum walk metric}\label{app:QWnon-sep}
\noindent
\textbf{Claim:} Given the non-unitary operation 
\begin{equation}
W(k)=e^{i\theta_1/2\sigma_1}e^{ik\sigma_3}e^{\gamma\sigma_3}e^{i\theta_2\sigma_1}e^{-\gamma\sigma_3}e^{ik\sigma_3}e^{i\theta_1/2\sigma_1}
\end{equation}
We define the walk operation for the total walk by $W=\sum_{k} \ket{k}\bra{k}\otimes W(k)$. For this operation, there does not exist any metric of the form $G=G_p\otimes G_c$ where $G_p$ is a metric in the momentum space and $G_c$ is a metric in the coin space.
\begin{proof}
The left eigenvectors (with the eigenvalues $\pm \epsilon(k)$) for $W(k)$ is given by
\begin{equation}\label{eigvec}
\begin{aligned}
	\ket{r_{\pm}}_k=&\begin{pmatrix}
	\pm(d_1+d_2) \\- d_3\pm \sin(\epsilon)
	\end{pmatrix} \\
	\text{where,}
	\\
	d_1=&\cosh (2 \gamma ) \cos (\theta_1) \sin (\theta_2)+\sin (\theta_1) \cos (\theta_2) \cos (2 k),\\
	d_2=&-\sin (\theta_2)\sinh (2 \gamma ),\\
	d_3=&\cos (\theta_2) \sin (2 k), \text{and} \\
	\epsilon_k=&\cos ^{-1}(\cos (\theta_1) \cos (\theta_2) \cos (2 k)\\
	&-\cosh (2 \gamma ) \sin (\theta_1) \sin (\theta_2))
\end{aligned}
\end{equation}
They are also the right eigenvectors of the operator $H^\dagger$ with eigenvalues $\epsilon^*_{\pm}$.
The metric for the complete quantum walk constructed with respect to the left eigenvectors of the Hamiltonian (represented by $G_0$) can be given by
\begin{equation}
\begin{aligned}
G_0=\sum_k \ket{k}\bra{k}\otimes G_0(k),\\
\text{where}\hspace{1cm} 
G_0(k)=\ket{r_+}_k\bra{r_+}+\ket{r_-}_k\bra{r_-}.
\end{aligned}
\end{equation}
Freedom in the choice of the metric: As discussed in the main text, if the Hamiltonian $H$ is hermitian under a metric $G$, then any other valid metric has the form $T^\dagger G^\sigma T$ \citep{Ali03}. We have constructed the $G_0^\sigma$ matrix as 
\begin{equation}
	G_0^\sigma=\sum_{k} \ket{k}\bra{k}\otimes \sigma_+(k) \ket{r_+}_k\bra{r_+}+\sigma_-(k)\ket{r_-}_k\bra{r_-}.
\end{equation}
where $\sigma_{\pm}(k)\in \{1,-1\}$. However, only if we choose the $\sigma_{\pm}(k)=1 \forall k$, we get a positive semi-definite metric. Positive semi-definiteness is required to ensure the norm of a state, defined by $\bra{\psi}G\ket{\psi}$, to be positive in $\mathcal{H}_G$. Therefore, the most general positive definite metric is of the form $T^\dagger G_0 T$ where $T$ commutes with the Hamiltonian $H$, or alternatively, $T^\dagger$ commutes with $H^\dagger$. Therefore, the action of $T^\dagger$ restricts the eigenvectors of $H^\dagger$ to the energy eigenspace of the $H^\dagger$, which in the case of the walk is 4-dimensional. In particular, if $k\in (0,\pi/2)$, this action is given by
\begin{equation}
\begin{aligned}
	 T^\dagger \ket{k}\ket{r(k)}&= a_0(k) \ket{k}\ket{r(k)}+a_1(k) \ket{-k}\ket{r(-k)}\\& 
	 +a_2(k)\ket{k-\pi}\ket{r(k-\pi)}+a_3(k) \ket{-k+\pi}\ket{r(k+\pi)}.
\end{aligned}
\end{equation}
Let us denote by $\ket{k^a}\ket{r_{\pm}(k^a)}$, where $a\in{0,1,2,3}$, the set of degenerate eigenvectors of $H^\dagger$ corresponding to the eigen energy $\pm \epsilon(k)$. We use the following assignment $k^0:=k\in (0,\pi
/2)$, $k^1:=-k$, $k^2:=k-\pi$ and $k^3:= -k+\pi/2$. Then the most general metric is given by
\begin{equation}\label{mostgenmetric}
\begin{aligned}
	G=\sum_{k}\sum_{a,b=0}^3 & \ket{k^a}\bra{k^b} \\
	& \otimes\Big( x_{ab}(k) \ket{r_+(k^a)}\bra{r_+(k^b)}+ y_{ab}(k)\ket{r_-(k^a)}_k\bra{r_-(k^b)}\Big).
\end{aligned}
\end{equation}
The Hermiticity requirement of the metric is given by the conditions $x_{ab}(k)^*=x_{ba}(k)$, and similarly for $y_{ab}(k)$, $\forall k$. The positive definiteness of the metric means that the $4\times 4$ matrices $x(k)$ and $y(k)$ have real and non-negative eigenvalues $\forall k$.
\\
We have therefore shown that the most general form of the metric for the 1-dimensional quantum walk in the momentum space is given as
\begin{equation}
	G=\sum_{k}\sum_{a,b=0}^3 \ket{k^a}\bra{k^b}\otimes G_{ab}(k).
\end{equation}
Hence, if $\exists$ a product form of the metric is given by $G_p\otimes G_c$, it must exist in the above form. In other words, we should be able to find appropriate functions $x_{ab}(y)$ and $y_{ab}(k)$, such that $G_{ab}(k)=f_{ab}(k) G_c$ $\forall k$ and $\forall a,b$, where $f_{ab}(k)=f_{ba}(k)*$ is a scalar function of $k$ and $G_c$ is a $k$-independent $2\times 2$ metric operator. While we should check this condition for all the 16 matrices $G_{ab}(k)$, we can make use of certain symmetries of the eigenspace to simplify the calculations. Firstly, we have $G_{ab}(k)=G_{ba}^\dagger (k)$. Second, we have $\ket{r^0(k)}=\ket{r^2(k)}$ and $\ket{r^1(k)}=\ket{r^3(k)}$. Therefore, to check whether the matrices $G_{ab}(k)$ can be written as $f_{ab}(k)G_c$, we need to check this condition only for $G_{00}(k)$, $G_{11}(k)$ and $G_{01}(k)$. Only if all of them can be made proportional to a $k$-independent matrix $G_c$, we can claim that the metric is compatible with the tensor product structure. Since we have to determine the $k$-independence of this matrix only up to a scalar function, we can take out the $x_{00}(k)$ factor. Therefore, we have to find $g_{00}\equiv x_{00}(k)/y_{00}(k)$, a bounded, real-valued function, such that $G_{00}(g_{00}(k))=( \ket{r_+}_k\bra{r_+}+g_{00}(k) \ket{r_-}_k\bra{r_-})$ is of the form $f_{00}(k) G_c$. 
\\
It will help us to introduce two variables to make the expressions more concise. Let $A=\dfrac{d_3(k^0)-\sin\epsilon((k^0))}{d_1(k^0)+d_2(k^0)}$ and $B=\dfrac{d_3(k^0)+\sin\epsilon(k^0)}{d_1(k^0)+d_2(k^0)}$. Then the three matrices can be written as 
\begin{equation}
\begin{aligned}
	&G_{00}(g_{00})=
		\left(
\begin{array}{cc}
g_{00}+1 &- (A+B g_{00}) \\
 -(A+Bg_{00}) & A^2+B^2g_{00}\\
\end{array}	
\right),\\
&G_{11}(g_{11})=
		\left(
\begin{array}{cc}
g_{11}+1 & B+A g_{11} \\
 B+Ag_{11} & B^2+A^2g_{11} \\
\end{array}	
\right),\\
&G_{01}(g_{01})=
		\left(
\begin{array}{cc}
g_{01}+1 & B+A g_{01} \\
 -(A+Bg_{01}) & BA(1+g_{01}) \\
\end{array}	
\right).
\end{aligned}
\end{equation}
If the $k$-dependence of the matrix $G_{ab}(g_{ab}(k))$ is contained in a factor $f_{ab}(k)$, the ratio of its any two components must be $k$-independent. For the case of $G_{00}$, we have the following two conditions
\begin{equation}
\begin{aligned}
	\dfrac{d}{dk}\Big( \dfrac{g_{00}B^2+A^2}{g_{00}+1}\Big)=0, \text{  and} \\	
\dfrac{d}{dk}\Big( \dfrac{g_{00}B+A}{y+1}\Big)=0.
\end{aligned}
\end{equation}
The two conditions translate to the following set of constraints on $g_{00}$
\begin{equation}
\begin{aligned}
\left(A^2-B^2\right) g_{00}'+2 (g_{00}+1) \left(BB'+A g_{00} A'\right)=0.\\
(A-B) g_{00}'+(g_{00}+1) \left(B'+g_{00}A'\right)=0.
\end{aligned}
\end{equation}
equating the $g_{00}'$ term in both Eq.s, we have $g_{00}=B'/A'$. However, this is a solution iff 
\begin{equation}
	g_{00}'=\Big( \dfrac{B'}{A'}\Big)'=^? \dfrac{2(B'+A')B'}{A'(A-B)}.
\end{equation}
The above relation is satisfied only if $\gamma=0$, in which case the metric turns out to be proportional to identity, as expected. The  calculations for the case of $G_{11}(k)$ are similar and we reach the conclusion that for $\gamma\neq 0$ there is no value of $g_{00}$ for which the matrix $G_{00}(k)$ is of the form $f_{00}(k)G_c$ (and similarly for $G_{11}(k)$ ).
For the case of $G_{01}(g_{01})$, the conclusion is straightforward: the independence of this matrix implies that $d(AB)/dk=0 \forall g_{01}(k)$. This can be verified to be not true.  
 Therefore, for $\gamma\neq 0$,  there does not exist a metric of the form $G_p\otimes G_c$ where $G_p$ and $G_c$ are positive definite metric operators.
\end{proof}
\section{Finding the intermediate map}\label{app:intermap}
In this section we construct the Choi matrix for the intermediate map. The map that takes a coin state (initially separable from the position space) from time $t=0$ to $t$ in the metric formalism is given by 
\begin{equation}
\begin{aligned}
	\Lambda(t,0)(\rho_c(t))&=Tr_p(W^t\rho_p\otimes\rho_c{W^\dagger}^t G(t))\\
	&=\dfrac{1}{2\pi}\int_{-\pi}^{\pi}W_c(k)^t \rho_c(t=0)(W_c(k)^\dagger)^t G(k,t)dk.
\end{aligned}
\end{equation}
In the normalized state method, the map will be defined without the metric. Unfortunately, this map does not have a nice analytical form in terms of the maps $\Lambda_k(t,t+1)$. We, therefore, derive it using vectorization.\\
Corresponding to the map, $\Lambda(t,0)$ we construct a matrix $L(t,0)$ such that
\begin{equation}
	L(t,0)=\{v_{11},v_{12},v_{21},v_{22}\},
\end{equation} 
where $v_{ij}=vec[\Lambda(t,0)\ket{i}\bra{j}]$. The composition of channels in the Eq. \eqref{mapcomp} translates to the multiplication of the corresponding matrix representations
\begin{equation}
	L(t+1,0)=L(t+1,t)L(t,0).
\end{equation}
This gives us the matrix representation of the intermediate map: $L(t+1,t)=L(t+1,0)L(t,0)^{-1}$. The Choi matrix of the intermediate map is constructed using the following operation
\begin{equation}
\mathcal{C}_{\Lambda}(t)=devec[U_{2\leftrightarrow 3}(L(t+1,t)\otimes I_4)U_{2\leftrightarrow 3}vec(\ket{\Phi}\bra{\Phi})],
\end{equation}
where $devec[]$ is the construction of the $4\times4$ matrix from the $16\times 1$ vector, and $U_{2\leftrightarrow 3}$ is the swap operator between the second and third subsystems
\begin{equation}
	U_{2\leftrightarrow 3}=I_2\otimes\begin{pmatrix}
	1&0&0&0\\
	0&0&1&0\\
	0&1&0&0\\
	0&0&0&1
	\end{pmatrix}\otimes I_2.
\end{equation}
The matrix $C_\Lambda(t)$ is exactly the Choi matrix of the intermediate map, and hence the function $g(t)$ from the Eq. \eqref{rhpdef} can be given in the following form for the discrete quantum walk evolution.
\end{document}